\documentclass[conference]{IEEEtran}
\usepackage{amsmath,amsfonts,amssymb}
\usepackage{array}
\usepackage{bm}
\usepackage{textcomp}
\usepackage{stfloats}
\usepackage{url}
\usepackage{verbatim}
\usepackage{graphicx}
\usepackage{cite}
\usepackage{color}
\usepackage{optidef}
\usepackage{physics}
\usepackage[caption=false,font=normalsize,labelfont=sf,textfont=sf]{subfig}
\usepackage{algorithm}
\usepackage{setspace}
\usepackage{algpseudocode}
\usepackage{etoolbox}
\usepackage{mathtools}

\newcommand\blfootnote[1]{%
  \begingroup
  \renewcommand\thefootnote{}\footnote{#1}%
  \addtocounter{footnote}{-1}%
  \endgroup
}

\begin{document}

\title{\huge Iterative Optimization of Reconfigurable Intelligent Surface Aided Single-Carrier Spatial Modulation}

\author{
\IEEEauthorblockN{Kentaro Nagaya and Shinya~Sugiura$^*$}
\IEEEauthorblockA{Institute of Industrial Science, The University of Tokyo\\
E-mail: \{nagaya-kentaro@g.ecc.u-tokyo.ac.jp, sugiura@iis.u-tokyo.ac.jp\} \vspace*{-5mm}
}

\thanks{The authors are with the Institute of Industrial Science, The University of Tokyo, Tokyo 153-8505, Japan (e-mail: sugiura@iis.u-tokyo.ac.jp). (\textit{Corresponding author: Shinya Sugiura}.)}
\thanks{This work was supported in part by the Japan Science and Technology Agency (JST) FOREST (Grant JPMJFR2127), in part by the JST ASPIRE (Grant JPMJAP2345), and in part by the Japan Society for the Promotion of Science (JSPS) KAKENHI (Grants 23H00470, 23K22752, 24K21615). Part of this paper was submitted for presentation at IEEE GLOBECOM 2025, Taipei, Taiwan~\cite{nagaya2025gc}.}
}

\markboth{}
{Shell \MakeLowercase{\textit{et al.}}: A Sample Article Using IEEEtran.cls for IEEE Journals}

\IEEEpubid{}

\maketitle

\begin{abstract}
This paper proposes a novel cyclic-prefixed single-carrier transmission scheme that amalgamate a reconfigurable intelligent surface (RIS) with spatial modulation in the frequency-selective fading channel. The discrete-input continuous-output memoryless channel's~(DCMC) capacities of the proposed schemes are formulated, while their gradients with respect to the RIS phase shifts are derived in the closed form. Then, the gradients are used for iteratively and efficiently optimizing the proposed RIS-aided schemes with the aid of the gradient-ascent algorithm. Our performance results demonstrate that the proposed iterative algorithm enhances the DCMC capacity of the schemes, while outperforming the conventional RIS benchmarks. Moreover, the convergence behavior and sensitivity in the proposed iterative algorithm are analyzed.
\end{abstract}


\section{Introduction}
\label{4 sec:Introduction}
\blfootnote{Preprint for publication in \textit{IEEE Global Communications Conference Workshop (GLOBECOM Wkshps)}, Taipei, Taiwan, 2025, DOI: 10.1109/GCWkshps68340.2025.11591036. $\copyright$ 2025 IEEE. Personal use of this material is permitted. Permission from IEEE must be obtained for all other uses, in any current or future media, including reprinting/republishing this material for advertising or promotional purposes, creating new collective works, for resale or redistribution to servers or lists, or reuse of any copyrighted component of this work in other works.}

The reconfigurable intelligent surface~(RIS)~\cite{wu2019intelligent} consists of planar reflective array elements, each having the capability of controlling the phase and amplitude to form desirable beam and polarization when reflecting an incoming incident wave. This enables the flexible optimization of channel environments so as to enhance the achievable communication performance. Furthermore, the RIS's benefits are attained in several scenarios, such as mmWave communication, a multi-cell downlink, power transfer to IoT sensors, a non-orthogonal multiple access system, a high-mobility cell-edge downlink, multicast transmission, and physical layer security. 
Index modulation~(IM) is an energy-efficient modulation technique that utilizes the indices in the space, time, and frequency dimensions to convey additional information~\cite{sugiura2017access}. Particularly when the indices of the antenna elements are used for IM, the technique is referred to as spatial modulation~(SM)~\cite{renzo2014pieee,sugiura2012comst}.

Importantly, the advent of the RIS concept opened new SM arrangements~\cite{basar2020reconfigurable,yan2020passive_wcl,yan2020passive_jsac,lin2021reconfigurable,guo2020reflecting,guo2021reconfigurable}.
In \cite{basar2020reconfigurable}, Basar merged the SM principle and the RIS device to configure the RIS-space shift keying and RIS-SM schemes in the frequency-flat channel, where the reflected beam to the index of the receive antennas is allowed to carry additional information bits.
In \cite{yan2020passive_wcl,yan2020passive_jsac}, Yan \textit{et al.}  the passive beamforming and information transfer~(PBIT) is introduced for the frequency-flat channel, where the RIS enhances the received power by reflective beamforming while additional information bits are modulated by activating and deactivating the RIS elements based on the SM principle. In \cite{lin2021reconfigurable}, Lin \textit{et al.} proposed a concept of reflection pattern modulation~(RPM) scheme for the frequency-flat channel, which exploits multiple RIS reflection patterns for the SM. In \cite{guo2020reflecting}, the reflecting modulation~(RM) scheme was proposed, which subsumes the conventional RIS transmission, the RIS-SM scheme, and the PBIT scheme by jointly mapping information bits onto the transmit signal and reflecting patterns, depending on the link arrangement between the transmitter and the RIS. Guo \textit{et al.} incorporated differential encoding into RM to dispense with coherent channel estimation~\cite{guo2021reconfigurable}.

Most of the previous studies of RIS-SM scheme assume the frequency-flat fading channel.
Also, while there are exceptional studies of RIS-aided SM schemes, which considered the frequency-selective fading channel~\cite{hidir2023on,yan2022frequency}, they assume the use of multi-carrier transmission, such as orthogonal frequency-division multiplexing (OFDM). However, as clarified in \cite{sugiura2015single,sugiura2017access}, the use of single-carrier transmission, rather than multi-carrier one, is essential to attain full performance benefits of SM since simultaneous switching of multiple analog antenna elements for different multicarriers is impractical. Naturally, the same single-carrier limitation is imposed on the RIS-SM scheme. To the best of our knowledge, there have been no RIS-SM studies considering wideband single-carrier transmission over the frequency-selective channel. Similarly, there has not been proposed any related RIS optimization algorithm for such realistic single-carrier RIS-SM scheme in the frequency-selective channel.

Against this background, the novel contributions of this paper are as follows. We propose novel single-carrier RIS-SM transmission scheme that operate in the frequency-selective fading channel. In the proposed scheme, SM is performed by activating a subset of the legitimate RIS groups. We formulate the accurate discrete-input continuous-output memoryless channel's~(DCMC) capacities~\cite{proakis2008digital} for the proposed schemes. Furthermore, the gradients of the DCMC capacity with respect to the RIS phase shift vector are derived in the closed form, which is then used for efficiently optimizing the proposed schemes with the aid of the gradient-ascent~(GA)-based algorithm.
It is demonstrated in our performance results that the proposed iterative algorithm enhances the DCMC capacity of the schemes, while outperforming the conventional RIS benchmarks. Moreover, the convergence behavior and sensitivity in the proposed iterative algorithm are analyzed in a comprehensive manner, which verifies its stability and efficiency.

\section{System Model}
\label{4 sec:System Model}
\subsection{Genaral Setup}
\label{4 subsec:General Setup}
In this paper, we consider RIS-aided single-carrier single-input-multiple-output~(SIMO) transmission between a single-antenna transmitter and a receiver that employs $R$ antennas, while the RIS is composed of $\it{M}$ reflection elements, similar to \cite{basar2020reconfigurable,yan2020passive_wcl,guo2021reconfigurable}. It is assumed that the transmitter and the RIS are connected via a stable fiber link, 
which allows accurate synchronization and cooperation.
A link between the transmitter and the receiver is assumed blocked, which necessitates the assistance of the RIS to establish a reliable connection. Note that in general, if a stable direct transmitter-receiver link exists, the assistance with the RIS may not be essential.

Furthermore, the RIS grouping~\cite{zheng2019intelligent} is employed for reducing calculation overhead imposed by a high $\it{M}$ value. More specifically, the RIS elements are divided into $\it{G}$ groups, each comprised of $\bar{\it{M}} = \it{M} / \it{G}$ adjoining elements.
Similar to \cite{zheng2019intelligent}, the RIS elements in each group are all assigned a single specific phase shift coefficient, which helps reduce the parameter size for optimization from $\it{M}$ to $\bar{\it{G}}$.

\subsection{Channel Model}
\label{4 subsec:Channel Model}
The frequency-selective fading channel is considered for each link between the transmitter, receiver, and the RIS. 
As above-mentioned, this assumption is necessary for practical RIS-aided SM schemes, while most previous studies considered the simplified frequency-flat channel.
Specifically, the channel coefficients between the transmitter and the RIS~(transmitter-RIS) and those between the RIS and receiver ~(RIS-receiver) are represented by $\bm{g}_m \in \mathbb{C}^{L_{1}}$ and
$\bm{h}_{r,m} \in \mathbb{C}^{L_{2}}  \ (m=1,\cdots,M, r=1,\cdots,R)$,
where $L_1$ and $L_2$ denote the number of delay taps, respectively.
The equivalent baseband channel impulse response (CIR) between the transmit antenna and the $r$-th receive antenna via the $m$-th RIS element~(transmitter-RIS-receiver) is defined as $\bm{g}_m * \bm{h}_{r,m}$, having the vector size of $L_{0} = L_{1} + L_{2} - 1$.

Let $\bm{\nu}_{r,m} = [(\bm{g}_m * \bm{h}_{r,m})^{T},\bm{0}_{1 \times (N-L_{0})}]^{T} \in \mathbb{C}^{N}$ denote the zero-padded transmitter-RIS-receiver CIR associated with the $r$-th receiver antenna and the $m$-th RIS element.
Furthermore, to represent the transmitter-RIS-receiver channel associated with the $g$-th RIS group,
let us define:
$\bar{\bm{\nu}}_{r,g} = \sum_{\tilde{m}}	\bm{\nu}_{r,m},~\tilde{m} \in \mathcal{I}_{g}$,
where $\mathcal{I}_{g} \triangleq \{i_{g,1}, \hdots, i_{g,\bar{M}}\}$ denote the indices of the RIS elements in the $g$-th RIS group.
Moreover, let us define the phase shifts for the $G$ RIS groups as $\bm{\theta} = [\theta_{1}, \hdots, \theta_{G}]^{T} \in \mathbb{C}^{G}$

\section{Single-Carrier RIS-Aided Spatial Modulation}
\label{4 sec:ON/OFF Spatial Modulation}

\subsection{Transmitter Model}
\label{4 subsec:SM-Transmitter Model}
At the transmitter, input bits are divided into two sets in each symbol duration.
For the $n$-th symbol, the first $\log_2K$ bits are modulated onto a complex-valued phase-shift keying (PSK) symbol $x_{n}$ from a $K$-sized constellation while
the second $\log_2J$ bits are mapped onto a single activated out of $J$ RIS group patterns, which is represented by $\bm{s}_{n} \in \{0,1\}^{G}$. Each entry of $\bm{s}_{n}$ represents the state of the corresponding RIS group, where $[\bm{s}_{n}]_{i}$ is associated with the $i$-th RIS group. To be more specific, when $[\bm{s}_{n}]_{i} = 0$, all the elements in the $i$-th RIS group are deactivated, implying that the associated RIS group does not reflect an impinging signal. By contrast, the $i$-th RIS group is activated when $[\bm{s}_{n}]_{i} = 1$, where the phase of the reflected signal is rotated by $\theta_{i}$.

Each transmission frame is composed of a data frame and a cyclic prefix (CP). The data frame includes $N_{b}$ PSK symbols $\bm{x} = [x_{1}, x_{2}, \hdots, x_{N_{b}}]^{T} \in \mathbb{C}^{N_{b}}$, each having the transmit power of $\sigma_{s}^{2}$. The CP corresponds to the last $N_{CP}$ symbols in the associated data frame. Note that we maintain $N_{CP}$ higher than the effective channel's delay tap length $L_{0}$ to avoid detrimental inter-block interference. Thus, the $n$-th PSK symbol of the overall transmission frame $\bar{\bm{x}} = [\bar{x}_1,\cdots,\bar{x}_{N_b+N_{CP}}]^T\in\mathbb{C}^{N_b+N_{CP}}$ is represented by
\begin{IEEEeqnarray}{rCL}
\label{4 equ: Transmission frame}
\bar{x}_n =
\left\{
\begin{array}{ll}
x_{n + N_{b} - N_{CP}}	&	n = 1,\hdots,N_{CP} \\
x_{n - N_{CP}} & n = N_{CP} + 1,\hdots,N_{CP} + N_{b}.
\end{array}
\right.
\end{IEEEeqnarray}
The deactivated RIS elements do not contribute to the signal reflections. Hence, to attain a high signal power reflected from the RIS, we only deactivate one RIS group for each RIS activation pattern, unlike the conventional SM-MIMO arrangement without an RIS~\cite{renzo2014pieee}, where all the power is transmitted from a single activated antenna element.

\subsection{Receiver Model}
\label{4 subsec:SM-Receiver Model}
At the receiver, after removing the samples associated with the CP from the received signal,
we obtain the $n$-th received samples $\bm{y}_{n} \in \mathbb{C}^{R}$ as follows:
$\bm{y}_{n} = \sum_{l=0}^{L_{0} - 1} x_{(n-l)_{N_{b}}} \mathbf{H}_{l} \mathbf{\Theta} \bm{s}_{(n-l)_{N_{b}}} + \bm{n}_{n}$,
where
$\mathbf{\Theta} = \textrm{diag}[\bm{\theta}] \in \mathbb{C}^{G \times G}$,
and $\mathbf{H}_{l} \in \mathbb{C}^{R \times G}$ is the CIR matrix corresponding to the $l$-th delay tap of the transmitter-RIS-receiver channel, whose $r$-th-row and $g$-th-column entry is given by
$[\mathbf{H}_{l}]_{r,g} = [\bar{\bm{\nu}}_{r,g}]_{l}$.
Also, $\bm{n}_{n} \in \mathbb{C}^{R}$ is the AWGN component, which follows the complex-valued Gaussian distribution of $\mathcal{N}_{c}(\mathbf{0}, \sigma_{n}^{2}\mathbf{I}_{N})$ and $\sigma_{n}^{2}$ represents the noise variance.
The overall received samples in a frame is expressed by
\begin{IEEEeqnarray}{rCL}
\bm{y} = \mathbf{H} (\mathbb{I}_{N_{b}} \otimes \mathbf{\Theta}) \mathbf{S} (\bm{x} \otimes \mathbf{1}_{G}) + \bm{n},
\label{eq:rx}
\end{IEEEeqnarray}
where
$\bm{y} = [\bm{y}_{1}^{T}, \bm{y}_{2}^{T},\hdots,\bm{y}_{N_{b}}^{T}]^{T} \in \mathbb{C}^{N_{r}N_{b}}$,
$\mathbf{S} = \textrm{diag}[\bm{s}_{1}^{T}, \bm{s}_{2}^{T}, \hdots, \bm{s}_{N_{b}}^{T}] \in \mathbb{C}^{G N_{b} \times G N_{b}}$, and
$\bm{n} = [\bm{n}_{1}^{T}, \bm{n}_{2}^{T},\hdots,\bm{n}_{N_{b}}^{T}]^{T} \in \mathbb{C}^{N_{r}N_{b}}$.
Moreover, $\mathbf{H}\in \mathbb{C}^{R N_{b} \times G N_{b}}$ is the block diagonal matrix consisting of the CIR matrices $\mathbf{H}_{l}$ as follows:
\begin{IEEEeqnarray}{lCL}
\mathbf{H} =
\begin{bmatrix}
\mathbf{H}_{0}&0&\hdots&\mathbf{H}_{L_{0}-1}&\mathbf{H}_{L_{0}-2}&\hdots&\mathbf{H}_{2}&\mathbf{H}_{1} \\
\mathbf{H}_{1}&\mathbf{H}_{0}&\hdots&0&\mathbf{H}_{L_{0}-1}&\hdots&\mathbf{H}_{3}&\mathbf{H}_{2} \\
\vdots&&&&&&&\vdots \\
0&0&\hdots&0&0&\hdots&\mathbf{H}_{1}&\mathbf{H}_{0} 
\end{bmatrix}. \nonumber
\end{IEEEeqnarray}
From \eqref{eq:rx}, the optimal maximum likelihood detection can be carried out for demodulating our RIS-aided SM symbols. Furthermore, to reduce the excessive detection complexity associated with the optimal detector, it is readily possible to employ the frequency-domain MMSE-based SM detector \cite{sugiura2015single}.
Note that in the rest of this paper, we focus our attention on the theoretical performance bounds limit, represented by average mutual information.

\subsection{Proposed RIS Optimization}
\label{4 subsec:SM-Proposed RIS Optimization}
In this section, we formulate the DCMC capacity of our single-carrier RIS-aided SM scheme. Then, the gradient vector of the DCMC capacity concerning the RIS phase shift vector is derived, which is used in the proposed GA-based RIS optimization algorithm.
The derivation of the proposed scheme's DCMC capacity allows us to evaluate the theoretical performance limit using discrete-valued complex signals and SM patterns, unlike the unconstrained capacity. Furthermore, it is exploited as the objective function for our RIS optimization, which is beneficial for accurately modeling the SM principle.

The DCMC capacity of the proposed scheme is expressed as~\cite{sugiura2010coherent}
\begin{IEEEeqnarray}{lLL}
\label{4 equ: mutual info}
&I(\bm{x}, \bm{s}; \bm{y}) 
= H(\bm{y}) - H(\bm{y} | \bm{s}, \bm{x}) \nonumber\\
&= \frac{1}{J^{N_{b}}} \frac{1}{K^{N_{b}}} \sum_{i=1}^{J^{N_{b}}} \sum_{j=1}^{K^{N_{b}}} 
\int_{\bm{y}} p(\bm{y} | \bm{s}_{i}, \bm{x}_{j}) \log_{2} \frac{p(\bm{y} | \bm{s}_{i}, \bm{x}_{j})}{p(\bm{y})}  d\bm{y}\nonumber\\
&= \frac{1}{J^{N_{b}}} \frac{1}{K^{N_{b}}} \sum_{i=1}^{J^{N_{b}}} \sum_{j=1}^{K^{N_{b}}} \int_{\bm{y}} p(\bm{y} | \bm{s}_{i}, \bm{x}_{j}) \nonumber\\
& \ \ \ \log_{2} \frac{p(\bm{y} | \bm{s}_{i}, \bm{x}_{j})}{\frac{1}{J^{N_{b}}} \frac{1}{K^{N_{b}}} \sum_{i^{'}=1}^{J^{N_{b}}} \sum_{j^{'}=1}^{K^{N_{b}}} p(\bm{y} | \bm{s}_{i^{'}}, \bm{x}_{j^{'}})}  d\bm{y}\nonumber\\
&= N_{b} \left( \log_{2} J + \log_{2} K \right) - \frac{1}{J^{N_{b}}} \frac{1}{K^{N_{b}}} \sum_{i=1}^{J^{N_{b}}} \sum_{j=1}^{K^{N_{b}}}\nonumber\\
& \ \ \ \int_{\bm{y}} p(\bm{y} | \bm{s}_{i}, \bm{x}_{j}) \log_{2} \frac{\sum_{i^{'}=1}^{J^{N_{b}}} \sum_{j^{'}=1}^{K^{N_{b}}} p(\bm{y} | \bm{s}_{i^{'}}, \bm{x}_{j^{'}})}{p(\bm{y} | \bm{s}_{i}, \bm{x}_{j})}  d\bm{y}.
\end{IEEEeqnarray}
The probability density function~(PDF) $p(\bm{y} | \bm{s}_{i}, \bm{x}_{j})$ is represented by
\begin{IEEEeqnarray}{rCL}
\label{4 equ: y conditional pdf}
p(\bm{y} | \bm{s}_{i}, \bm{x}_{j}) = \left( \frac{1}{\pi \sigma_{n}^{2}} \right)^{N_{b} R} \exp(- \frac{\left\| \bm{y}  - \bm{f}_{\bm{s}_{i}, \bm{x}_{j}} \right\|^{2}}{\sigma_{n}^{2}}),
\end{IEEEeqnarray}
where
$\bm{f}_{\bm{s}_{i}, \bm{x}_{j}} = \mathbf{H} (\mathbb{I}_{N_b} \otimes \mathbf{\Theta}) \mathbf{S}_{i} (\bm{x}_{j} \otimes \mathbf{1}_{G})$.
Furthermore, with \eqref{4 equ: y conditional pdf}, the logarithm of \eqref{4 equ: mutual info} is simplified to
{\small
\begin{IEEEeqnarray}{lCr}
\log_2\frac{\sum_{i^{'}=1}^{J^{N_{b}}} \sum_{j^{'}=1}^{K^{N_{b}}} p(\bm{y} | \bm{s}_{i^{'}}, \bm{x}_{j^{'}})}{p(\bm{y} | \bm{s}_{i}, \bm{x}_{j})}= \log_2\sum_{i^{'}=1}^{J^{N_{b}}} \sum_{j^{'}=1}^{K^{N_{b}}} \exp(\Psi_{i^{'},j^{'}}^{i,j}), \nonumber \\ \label{4 equ: expression inside the logarithm}
\end{IEEEeqnarray}}
where
\begin{IEEEeqnarray}{lCr}
\label{4 equ: Psi}
\Psi_{i^{'},j^{'}}^{i,j} = \frac{\left\| \bm{n}\right\|^2 - \left\| \bm{f}_{\bm{s}_{i}, \bm{x}_{j}} - \bm{f}_{\bm{s}_{i^{'}}, \bm{x}_{j^{'}}} + \bm{n}\right\|^2}{\sigma_{n}^{2}}.
\end{IEEEeqnarray}
Inserting \eqref{4 equ: expression inside the logarithm} into \eqref{4 equ: mutual info} yields
\begin{IEEEeqnarray}{rCL}
\label{4 equ: mutual info 2}
&I(\bm{x}, \bm{s}; \bm{y}) = 
N_{b} \left( \log_{2} J + \log_{2} K \right) - \frac{1}{J^{N_{b}}} \frac{1}{K^{N_{b}}} \nonumber\\
& \ \ \ \sum_{i=1}^{J^{N_{b}}} \sum_{j=1}^{K^{N_{b}}} \mathbb{E}_{\bm{y} | \bm{s}_{i}, \bm{x}_{j}} \left[ \log_{2} \left[\sum_{i^{'} = 1}^{J^{N_{b}}} \sum_{j^{'} = 1}^{K^{N_{b}}}\exp(\Psi_{i^{'}, j^{'}}^{i,j}) \right] \right], \ \ \
\end{IEEEeqnarray}
noting that the integral notation is switched to expectation operation for ease of exposition.

Based on the DCMC capacity derived above, we formulate a maximization problem to attain the optimal RIS phase shifts as follows:
\begin{maxi!}|s|
{\mathbf{\Theta}}
{(\ref{4 equ: mutual info 2})}
{\label{4 opt:mutual info 1}}
{}
\addConstraint{|\theta_{n}|^{2}}{= 1, \forall n \in 1,\hdots,G}.
\end{maxi!}
However, the optimization over the entire data frame requires substantial computations since the two-fold summation $\sum_{i=1}^{J^{N_{b}}} \sum_{j=1}^{K^{N_{b}}}$ imposes complexity order of $\order{J^{N_{b}} K^{N_{b}}}$. Hence, the optimization problem (\ref{4 opt:mutual info 1}) is modified to reduce the frame length while maintaining the characteristics of the channel with a delay tap of $L_{0}$. With the removal of the constant terms, the optimization problem (\ref{4 opt:mutual info 1}) becomes
\begin{maxi!}|s|
{\mathbf{\Theta}}
{-\sum_{i=1}^{J^{L_{0}}} \sum_{j=1}^{K^{L_{0}}} \mathbb{E}_{\bm{y} | \bm{s}_{i}, \bm{x}_{j}} \left[ \log_{2} \left[\sum_{i^{'} = 1}^{J^{L_{0}}} \sum_{j^{'} = 1}^{K^{L_{0}}}\exp(\Psi_{i^{'}, j^{'}}^{i,j}) \right] \right]}
{\label{4 opt:mutual info 2}}
{}
\addConstraint{|\theta_{n}|^{2}}{= 1, \forall n \in 1,\hdots,G},
\end{maxi!}
where the number of summations are reduced from $(J^{N_{b}}K^{N_{b}})^2$ to $(J^{L_{0}}K^{L_{0}})^2$.

Our algorithm employs the GA approach to solve the optimization problem (\ref{4 opt:mutual info 2}).
Let us consider the real-valued RIS phase shift vector of $\boldsymbol{\phi} = [\phi_{1}, \phi_{2}, \hdots, \phi_{G}]^T \in \mathbb{R}^G$, having the relationship of $\theta_g = e^{j\theta_g}$.
Then, the gradient of the DCMC capacity concerning the RIS phase shift vector $\boldsymbol{\phi} $ is derived as \eqref{4 equ: gradient}, which is shown on the top of the next page.
\begin{figure*}
{\small
\begin{IEEEeqnarray}{lL}
\label{4 equ: gradient}
\nabla_{\bm{\phi}} I(\bm{x},\bm{s};\bm{y}) &= - \sum_{i=1}^{J^{L_{0}}} \sum_{j=1}^{K^{L_{0}}} \nabla_{\bm{\phi}} \mathbb{E}_{\bm{y}|\bm{s}_{i},\bm{x}_{j}} \left[ \log_{2} \left[ \sum_{i^{'}=1}^{J^{L_{0}}} \sum_{j^{'}=1}^{K^{L_{0}}} \exp(\Psi_{i^{'},j^{'}}^{i,j}) \right]\right]\nonumber\\
&= - \sum_{i=1}^{J^{L_{0}}} \sum_{j=1}^{K^{L_{0}}} \mathbb{E}_{\bm{y}|\bm{s}_{i},\bm{x}_{j}} \left[ \frac{1}{\sum_{i^{'}=1}^{J^{L_{0}}} \sum_{j^{'}=1}^{K^{L_{0}}} \exp(\Psi_{i^{'},j^{'}}^{i,j})} \frac{1}{\ln(2)} \left[ \sum_{i^{'}=1}^{J^{L_{0}}} \sum_{j^{'}=1}^{K^{L_{0}}} \exp(\Psi_{i^{'},j^{'}}^{i,j}) \nabla_{\bm{\phi}} \Psi_{i^{'},j^{'}}^{i,j} \right]\right].
\end{IEEEeqnarray}
\hrulefill
}
\end{figure*}
\begin{figure*}
{\small
\begin{IEEEeqnarray}{lL}
\label{4 equ: grad Psi}
\nabla_{\bm{\phi}} \Psi_{i^{'},j^{'}}^{i,j} &= \nabla_{\bm{\phi}} \frac{\left\| \bm{n} \right\|^2 - \left\| \mathbf{H} \mathbf{S}_{i} \left( \mathbf{X}_{j} \otimes \mathbb{I}_{G} \right) \left( \bm{1}_{L_{0}} \otimes \bm{\theta} \right) - \mathbf{H} \mathbf{S}_{i^{'}} \left( \mathbf{X}_{j^{'}} \otimes \mathbb{I}_{G} \right) \left( \bm{1}_{L_{0}} \otimes \bm{\theta} \right) + \bm{n}\right\|^2}{\sigma_{n}^{2}}\nonumber\\
&= - \frac{1}{\sigma_{n}^{2}} \nabla_{\bm{\phi}} \left\| \mathbf{H} \mathbf{T}_{i^{'},j^{'}}^{i,j} \tilde{\bm{\theta}} + \bm{n}\right\|^2.
\end{IEEEeqnarray}
\hrulefill
}
\end{figure*}
In order to express $\nabla_{\bm{\phi}} \Psi_{i^{'},j^{'}}^{i,j}$ 
in the closed form,
let us transform $\bm{f}_{\bm{s}_{i}, \bm{x}_{j}}$ into a more tractable form as follows:
\begin{IEEEeqnarray}{rCL}
\label{4 equ: f tractable form}
\bm{f}_{\bm{s}_{i}, \bm{x}_{j}} = \mathbf{H} \mathbf{S}_{i} \left( \mathbf{X}_{j} \otimes \mathbb{I}_{G}\right)	\left( \mathbf{1}_{L_{0}} \otimes \bm{\theta} \right),
\end{IEEEeqnarray}
where $\mathbf{X}_j=\mathrm{diag}\left\{\bm{x}_j\right\}$.
By substituting (\ref{4 equ: f tractable form}) into (\ref{4 equ: Psi}),
we arrive at the gradient of $\nabla_{\bm{\phi}} \Psi_{i^{'},j^{'}}^{i,j}$ as \eqref{4 equ: grad Psi}, shown on the top of the next page, where
$\mathbf{T}_{i^{'},j^{'}}^{i,j} = \mathbf{S}_{i} \left(\mathbf{X}_{j} \otimes \mathbb{I}_{G}\right) - \mathbf{S}_{i^{'}} (\mathbf{X}_{j^{'}} \otimes \mathbb{I}_{G})$ and
$\tilde{\bm{\theta}} = \bm{1}_{L_{0}} \otimes \bm{\theta}$.

Furthermore, (\ref{4 equ: grad Psi}) is transformed to:~\cite{brandwood1983complex}
{\small
\begin{IEEEeqnarray}{lL}
\label{4 equ: grad Psi 2}
\nabla_{\bm{\phi}} \Psi_{i^{'},j^{'}}^{i,j} &=
- \frac{1}{\sigma_{n}^{2}} \nabla_{\bm{\phi}} \left[ \left( \tilde{\bm{\theta}}^{H}\left(\mathbf{T}_{i^{'},j^{'}}^{i,j}\right)^{H} \mathbf{H}^{H} + \bm{n}^{H} \right) \right.
\nonumber\\
&\left. \left( \tilde{\bm{\theta}} \mathbf{H} \mathbf{T}_{i^{'},j^{'}}^{i,j} + \bm{n}\right) \right]
\nonumber\\
&= - \frac{1}{\sigma_{n}^{2}} \nabla_{\bm{\phi}} \left[ \tilde{\bm{\theta}}^{H} \tilde{\mathbf{P}}_{i^{'},j^{'}}^{i,j} \tilde{\bm{\theta}} + \tilde{\bm{\theta}}^{H} \tilde{\bm{p}}_{i^{'},j^{'}}^{i,j} + \left(\tilde{\bm{p}}_{i^{'},j^{'}}^{i,j}\right)^{H} \tilde{\bm{\theta}}\right] \nonumber\\
&= - \frac{1}{\sigma_{n}^{2}} \nabla_{\bm{\phi}} \left[ \bm{\theta}^{H} \mathbf{P}_{i^{'},j^{'}}^{i,j} \bm{\theta} + \bm{\theta}^{H} \bm{p}_{i^{'},j^{'}}^{i,j} + \left(\bm{p}_{i^{'},j^{'}}^{i,j}\right)^{H} \bm{\theta}\right], \nonumber\\
\end{IEEEeqnarray}
}
where
$\tilde{\mathbf{P}}_{i^{'},j^{'}}^{i,j} =  \left( \mathbf{T}_{i^{'},j^{'}}^{i,j}\right)^{H}\mathbf{H}^{H} \mathbf{H} \mathbf{T}_{i^{'},j^{'}}^{i,j}$ and
$\tilde{\bm{p}}_{i^{'},j^{'}}^{i,j} = \left( \mathbf{T}_{i^{'},j^{'}}^{i,j}\right)^{H}\mathbf{H}^{H} \bm{n}$.
The transition from the second line to the third line of (\ref{4 equ: grad Psi 2}) comes from the transformation of the matrix $\tilde{\mathbf{P}}_{i^{'},j^{'}}^{i,j}$ and the vector $\tilde{\bm{p}}_{i^{'},j^{'}}^{i,j}$ as
$\left[\mathbf{P}_{i^{'},j^{'}}^{i,j}\right]_{k,l} = \sum_{m \in \mathcal{M}_{k}} \sum_{n \in \mathcal{M}_{l}} \left[\tilde{\mathbf{P}}_{i^{'},j^{'}}^{i,j}\right]_{m,n}$ and 
$\left[ \bm{p}_{i^{'}, j^{'}}^{i,j} \right]_{k} = \sum_{m \in \mathcal{M}_{k}} \left[ \tilde{\bm{p}}_{i^{'}, j^{'}}^{i,j} \right]_{m}$ for
$\mathcal{M}_{i} \triangleq \{i, G + i, \hdots, (L_{0} - 1)G + i\}, k, l \in \{1, \dots, G\}$.
Since $\Psi_{i^{'},j^{'}}^{i,j}$ may be the function of the complex variables $\{\theta_{1}, \theta_{2}, \hdots, \theta_{G}, \theta_{1}^{*}, \theta_{2}^{*}, \hdots, \theta_{G}^{*}\}$,
the partial derivative of $\Psi_{i^{'},j^{'}}^{i,j}$ with respect to the real variable $\phi_{g}$ is expressed as
\begin{IEEEeqnarray}{rCL}
\label{4 equ: partial derivative start}
\frac{\partial \Psi_{i^{'},j^{'}}^{i,j}}{\partial \phi_{g}}  &=& \sum_{m=1}^{G} \frac{\partial \Psi_{i^{'},j^{'}}^{i,j}}{\partial \theta_{m}} \frac{\partial \theta_{m}}{\partial \phi_{g}} + \frac{\partial \Psi_{i^{'},j^{'}}^{i,j}}{\partial \theta_{m}^{*}} \frac{\partial \theta_{m}^{*}}{\partial \phi_{g}} \\
\label{4 equ: partial derivative end}
&=& j \theta_{g} \frac{\partial \Psi_{i^{'},j^{'}}^{i,j}}{\partial \theta_{g}}  -j \theta_{g}^{*} \frac{\partial \Psi_{i^{'},j^{'}}^{i,j}}{\partial \theta_{g}^{*}}. 
\end{IEEEeqnarray}
Therefore, $\nabla_{\bm{\phi}} \Psi_{i^{'},j^{'}}^{i,j}$ may be represented by
\begin{equation}
\label{4 equ: grad Psi 3}
\nabla_{\bm{\phi}} \Psi_{i^{'},j^{'}}^{i,j} = j \mathbf{\Theta}\nabla_{\bm{\theta}} \Psi_{i^{'},j^{'}}^{i,j} - j \mathbf{\Theta}^{H} \nabla_{\bm{\theta}^{*}} \Psi_{i^{'},j^{'}}^{i,j},
\end{equation}
where
\begin{IEEEeqnarray}{rCL}
\label{4 equ: nabla theta}
\nabla_{\bm{\theta}} &=& \left[\frac{\partial}{\partial \theta_{1}}, \frac{\partial}{\partial \theta_{2}}, \hdots, \frac{\partial}{\partial \theta_{G}}\right]^{T} \\
\label{4 equ: nabla theta conj}
\nabla_{\bm{\theta}^{*}} &=& \left[\frac{\partial}{\partial \theta_{1}^{*}}, \frac{\partial}{\partial \theta_{2}^{*}}, \hdots, \frac{\partial}{\partial \theta_{G}^{*}}\right]^{T}.
\end{IEEEeqnarray}
Applying \eqref{4 equ: nabla theta} and \eqref{4 equ: nabla theta conj} to \eqref{4 equ: Psi} yields the gradients of $\Psi_{i^{'},j^{'}}^{i,j}$ with respect to the complex vectors $\bm{\theta}$ and $\bm{\theta}^{*}$ as~\cite{brandwood1983complex}
\begin{equation}
\begin{split}
&\nabla_{\bm{\theta}} \Psi_{i^{'},j^{'}}^{i,j} = - \frac{1}{\sigma_{n}^{2}} \left( \left(\mathbf{P}_{i^{'},j^{'}}^{i,j}\right)^{*} \bm{\theta}^{*} + \left(\bm{p}_{i^{'},j^{'}}^{i,j}\right)^{*} \right) \\
&\nabla_{\bm{\theta}^{*}} \Psi_{i^{'},j^{'}}^{i,j} = - \frac{1}{\sigma_{n}^{2}} \left( \mathbf{P}_{i^{'},j^{'}}^{i,j} \bm{\theta} + \bm{p}_{i^{'},j^{'}}^{i,j} \right).
\end{split}
\end{equation}
Note that $\nabla_{\bm{\theta}} \Psi_{i^{'},j^{'}}^{i,j}$ and $\nabla_{\bm{\theta}^{*}} \Psi_{i^{'},j^{'}}^{i,j}$ have the conjugate relationship.

Thus, (\ref{4 equ: grad Psi 3}) becomes
\begin{IEEEeqnarray}{lL}
\label{4 equ: grad Psi 4}
\nabla_{\bm{\phi}} \Psi_{i^{'},j^{'}}^{i,j} =&-2 \frac{1}{\sigma_{n}^{2}} \Im \left( \mathbf{\Theta}^{H} \left( \mathbf{P}_{i^{'},j^{'}}^{i,j} \bm{\theta} + \bm{p}_{i^{'},j^{'}}^{i,j} \right) \right).
\end{IEEEeqnarray}
The substitution of (\ref{4 equ: grad Psi 4}) into (\ref{4 equ: gradient}) allows us to arrive at the gradients needed for the optimization problem~(\ref{4 opt:mutual info 2}).

\subsection{Performance Results}
\label{4 subsec:SM-Performance Results}
We carried out Monte Carlo simulations to evaluate the achievable performance of the proposed RIS optimization algorithm.
It is assumed that the transmitter, receiver, and RIS are located at $(x, y, z) = (5, 150, 1)$, $(150, 0, 2)$, and $(0, 150, 2)$, respectively. The single-slope model $P_{r}/P_{t} = K(d/d_{r})^{\gamma}$ was used to calculate the path loss, where $P_{t}$ and $P_{r}$ denote the transmit and received powers at transmitter and receiver, respectively, while we considered $d_{r} = 1$~m and $K=-30$~dB.

The frequency-selective Rician fading was used for generating the channel coefficients, where the average power of each link was fixed to that calculated by the path-loss model. Also, the non-LoS (NLoS) components were generated based on a uniform power delay profile. The numbers of the delay taps were set as $L_{1} = 1$ and $L_{2} = 2$, and hence, the overall end-to-end delay taps was $L_{0} = L_{1} + L_{2} - 1 = 2$. The Rician $K$ factor for the RIS-receiver and transmitter-RIS links were set to $3$~dB and $0$~dB, respectively.
The number of the RIS elements and that of the RIS groups were given by $M = 64$ and $G = 16$, while the receiver was equipped with $R=4$ antennas.

In our simulations, the initial RIS phase shifts were generated randomly according to the uniform distribution [$0$, $2\pi$].
We considered the constellation size of $K=4$ and the legitimate RIS group patterns of $J=4$, hence having $4$~bpcu. This implies that in each symbol duration, the transmitter sends a QPSK symbol while activating one of the four RIS group patterns, where in each RIS group pattern, one of the four RIS groups is turned off. The step-size and convergence threshold were maintained to $\mu=1$ and $\epsilon = 1 \times 10^{-7}$, unless othersize noted.

\subsubsection{Convergence Analysis}
\label{4 subsubsec:SM-Convergence Analysis}
\begin{figure}[!t]
\centering
\subfloat[]{\includegraphics[width=0.25\linewidth]{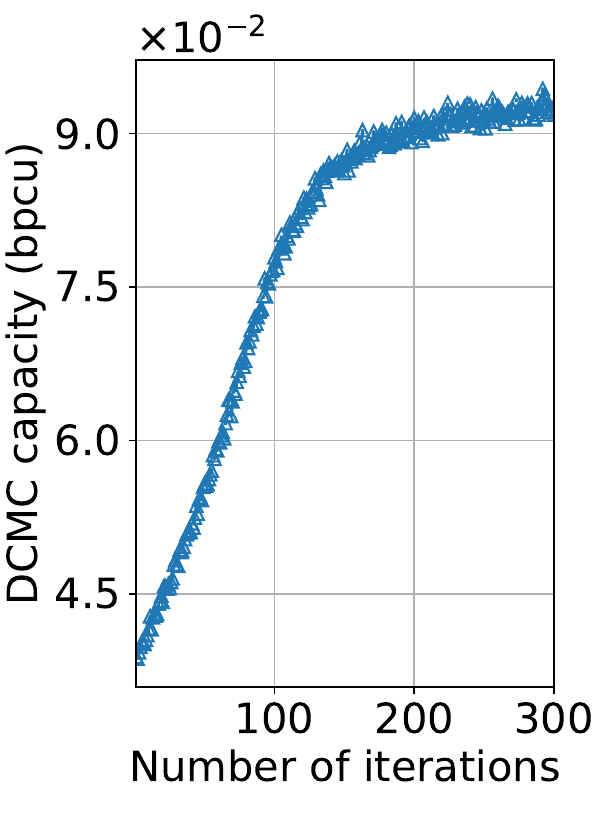}%
\label{4 fig:RIS opt Power T 0dBm}}
\hfill
\subfloat[]{\includegraphics[width=0.25\linewidth]{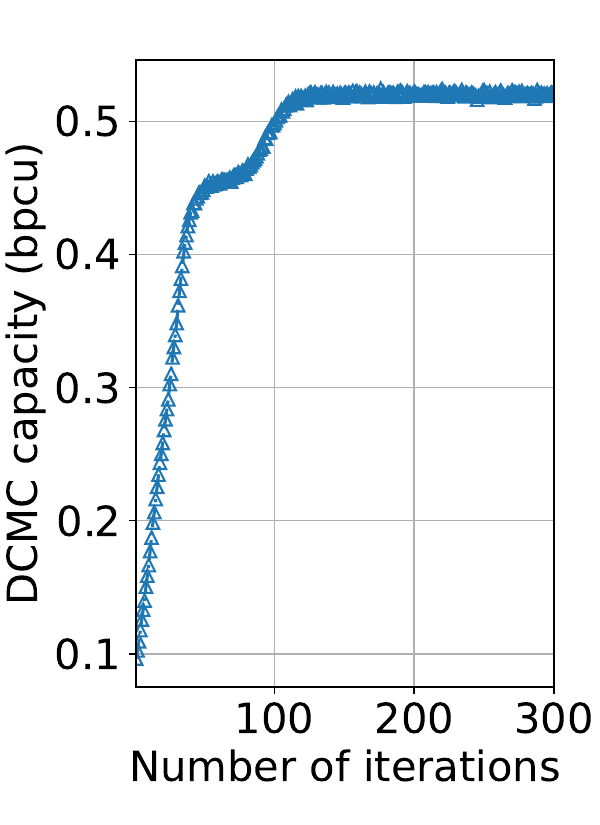}%
\label{4 fig:RIS opt Power T 10dBm}}
\subfloat[]{\includegraphics[width=0.25\linewidth]{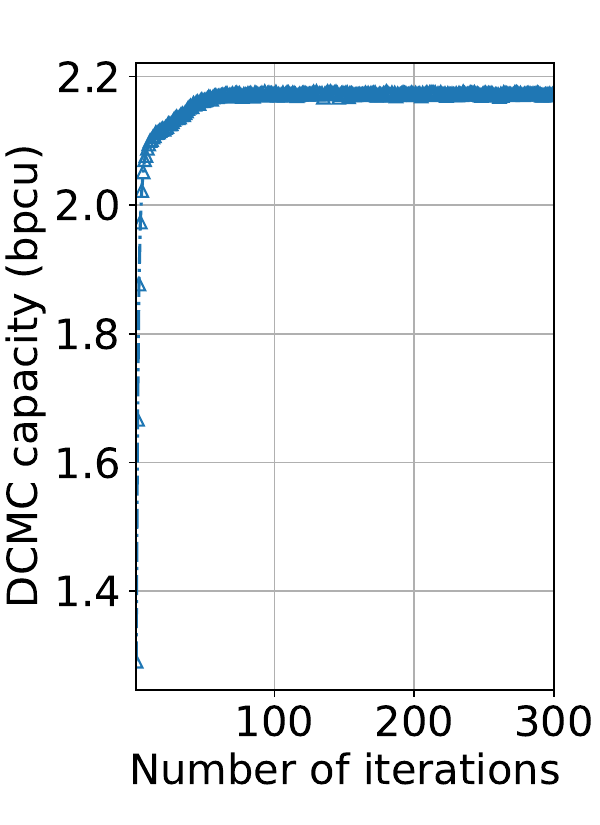}%
\label{4 fig:RIS opt Power T 20dBm}}
\hfill
\subfloat[]{\includegraphics[width=0.25\linewidth]{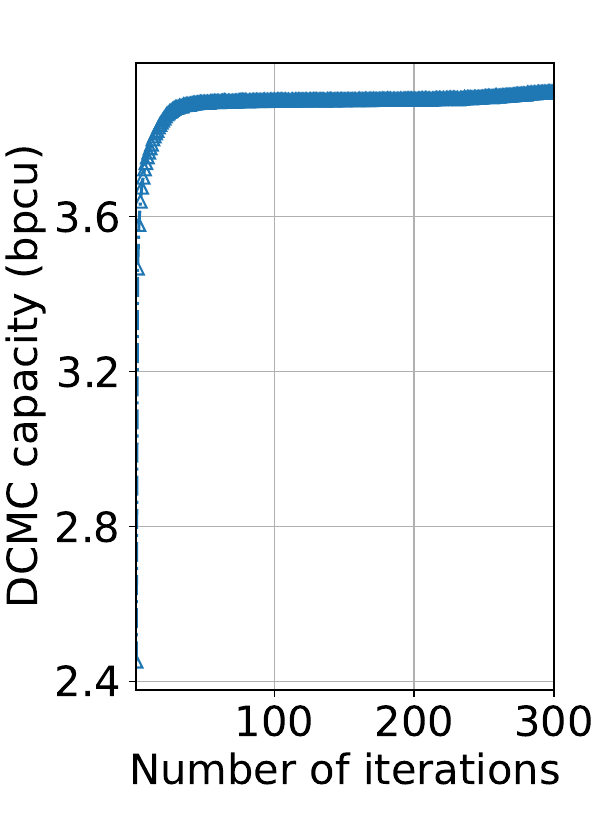}%
\label{4 fig:RIS opt Power T 30dBm}}
\caption{The convergence behavior of the DCMC capacity in our iterative optimization algorithm: (a) $P_{t} = 0$~dBm, (b) $P_{t} = 10$~dBm, (c) $P_{t} = 20$~dBm, and (d) $P_{t} = 30$~dBm.}
\label{4 fig:RIS opt Power T}
\end{figure}
We evaluated the convergence behavior of the proposed GA-based RIS optimization algorithm.
For the comparative convergence analysis, the convergence threshold was set to $\epsilon=0$.
Fig.~\ref{4 fig:RIS opt Power T} shows the convergence behavior of the DCMC capacity in the proposed algorithm.
The transmit power is set to $P_{t}=0$~dBm, $10$~dBm, $20$~dBm, and $30$~dBm in Figs.~\ref{4 fig:RIS opt Power T 0dBm}, \ref{4 fig:RIS opt Power T 10dBm}, \ref{4 fig:RIS opt Power T 20dBm}, and \ref{4 fig:RIS opt Power T 30dBm}, respectively.
Observed in Fig.~\ref{4 fig:RIS opt Power T} that in each transmit power scenario, the proposed algorithm increases the DCMC capacity upon increasing the number of iterations until the convergence. Also, it was seen that for the higher transmit power scenario, the algorithm tended to converge more quickly.
This is because the absolute value of the gradient \eqref{4 equ: gradient} depends on the transmit power, i.e., a higher step size tends to be required for a lower transmit power.
Note that the rattling of the DCMC capacity was caused due to the expectation operation $\mathbb{E}_{\bm{y} | \bm{s}_{i}, \bm{x}_{j}} \left[ \cdot \right]$ in \eqref{4 equ: mutual info 2}, whose calculations imposes an extensive amount of samplings while the effects AWGN components are relatively high compared to the transmit power.

\begin{figure}[!t]
\centering
\subfloat[]{\includegraphics[width=0.25\linewidth]{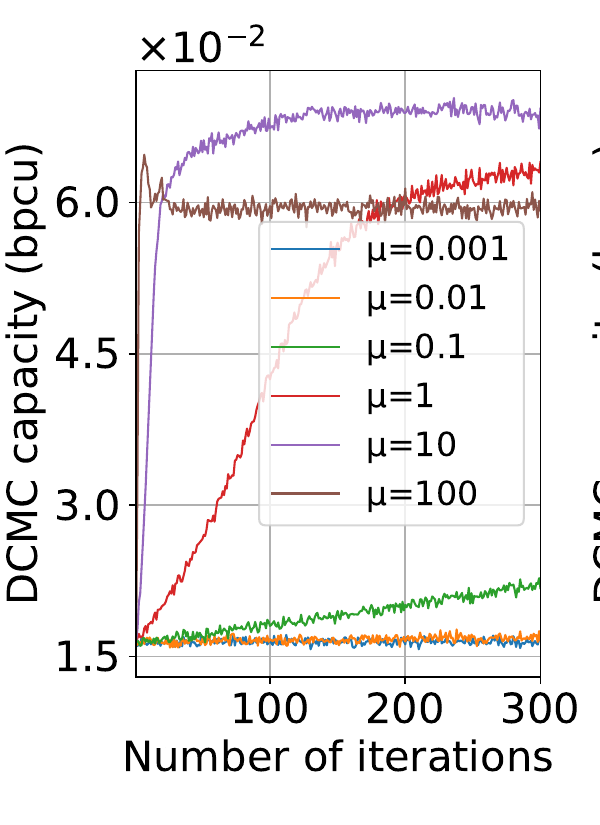}%
\label{4 fig:RIS opt Power T vary mu 0dBm}}
\hfill
\subfloat[]{\includegraphics[width=0.25\linewidth]{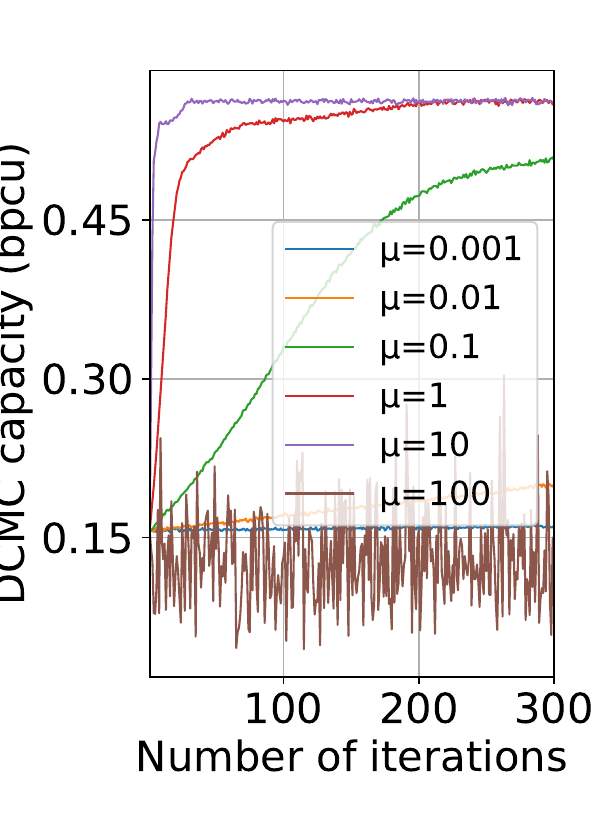}%
\label{4 fig:RIS opt Power T vary mu 10dBm}}
\subfloat[]{\includegraphics[width=0.25\linewidth]{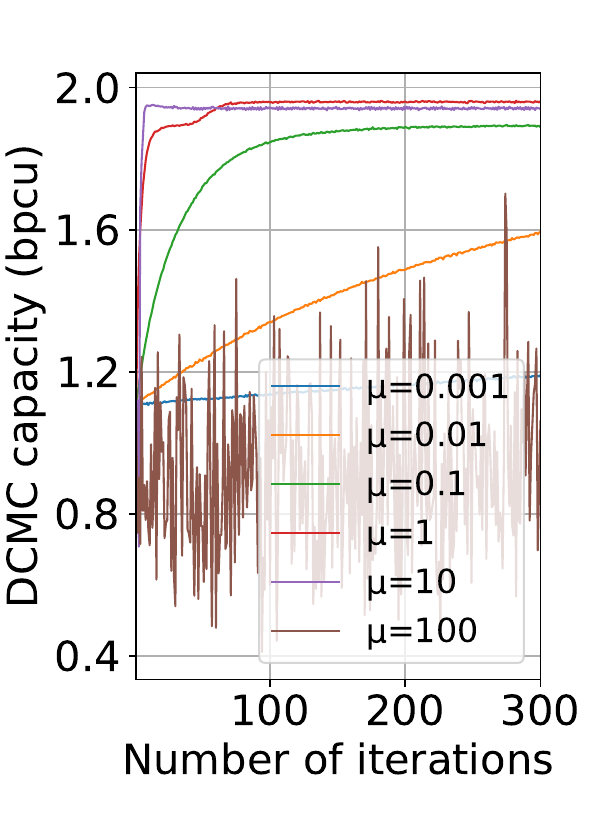}%
\label{4 fig:RIS opt Power T vary mu 20dBm}}
\hfill
\subfloat[]{\includegraphics[width=0.25\linewidth]{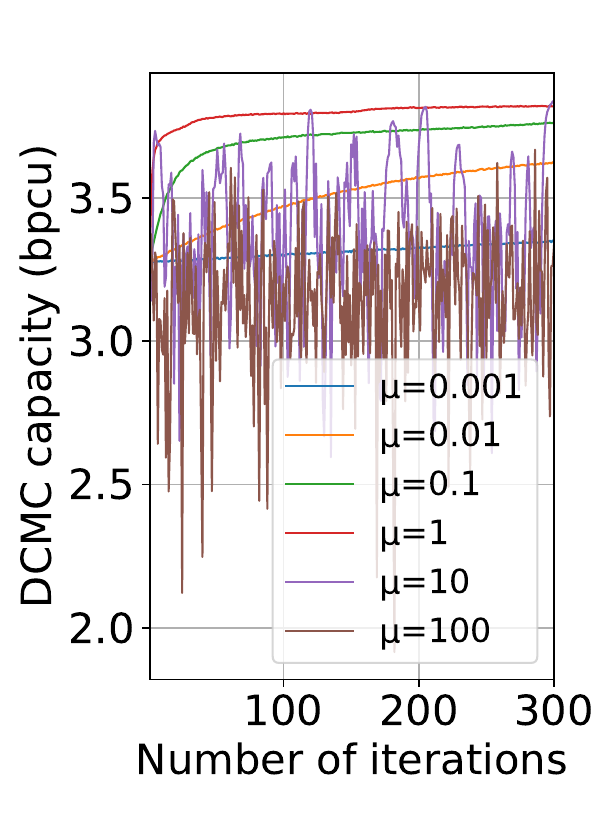}%
\label{4 fig:RIS opt Power T vary mu 30dBm}}
\caption{Convergence analysis of the DCMC capacity for the varied number of iterations from $\mu=0.001$ to $100$: (a) $P_{t} = 0$~dBm, (b) $P_{t} = 10$~dBm, (c) $P_{t} = 20$~dBm, and (d) $P_{t} = 30$~dBm.}
\label{4 fig:RIS opt Power T vary mu}
\end{figure}

\begin{figure}[!t]
\centering
\subfloat[]{\includegraphics[width=0.25\linewidth]{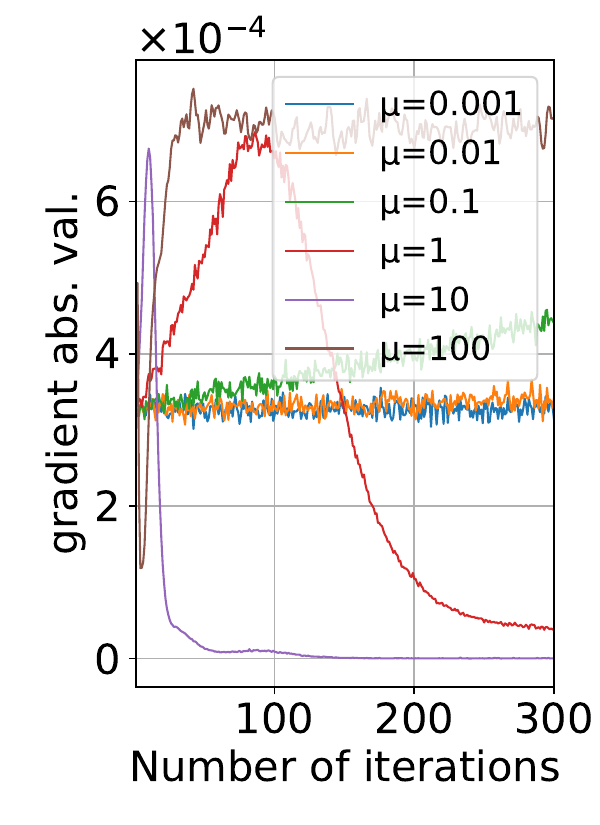}%
\label{4 fig:RIS grad abs Power T vary mu 0dBm}}
\hfill
\subfloat[]{\includegraphics[width=0.25\linewidth]{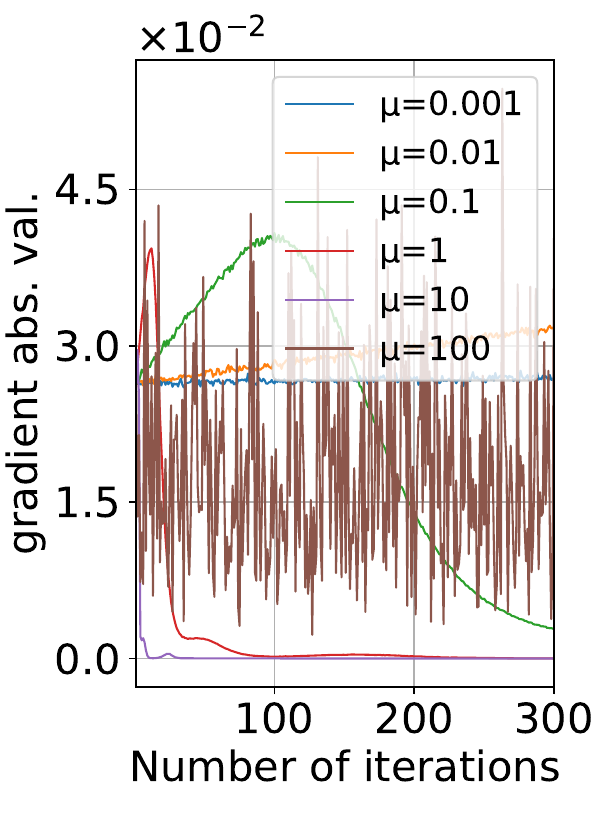}%
\label{4 fig:RIS grad abs Power T vary mu 10dBm}}
\subfloat[]{\includegraphics[width=0.25\linewidth]{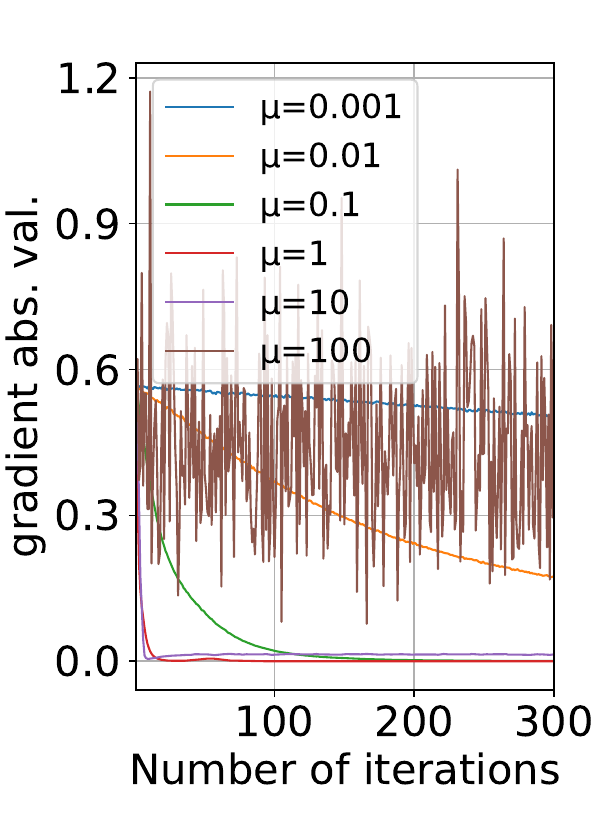}%
\label{4 fig:RIS grad abs Power T vary mu 20dBm}}
\hfill
\subfloat[]{\includegraphics[width=0.25\linewidth]{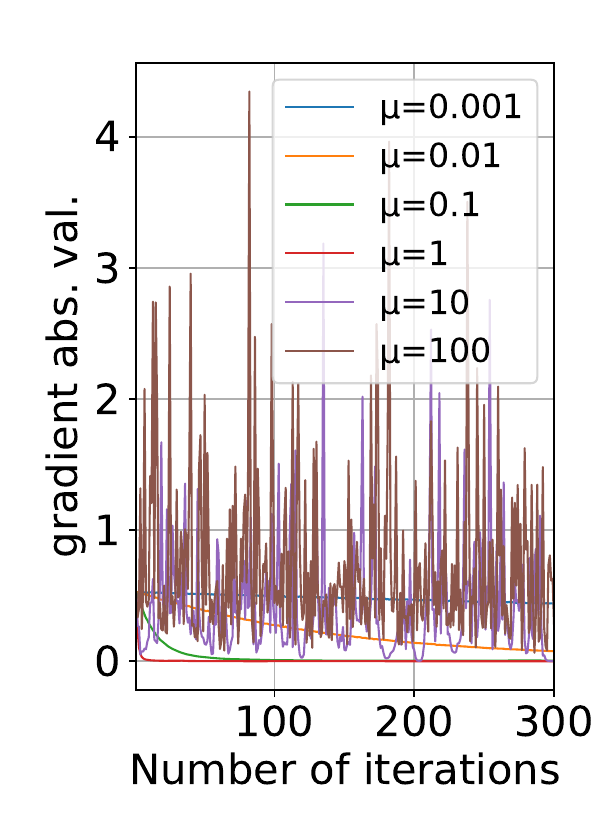}%
\label{4 fig:RIS grad abs Power T vary mu 30dBm}}
\caption{Convergence analysis of $\|\nabla_{\bm{\phi}} \Psi_{i^{'},j^{'}}^{i,j}\|$ for the varied number of iterations from $\mu=0.001$ to $100$: (a) $P_{t} = 0$~dBm, (b) $P_{t} = 10$~dBm, (c) $P_{t} = 20$~dBm, and (d) $P_{t} = 30$~dBm.}
\label{4 fig:RIS grad abs Power T vary mu}
\end{figure}

Next, we conducted the sensitivity analysis of the proposed optimization scheme in terms of the step size $\mu$.
Figs.~\ref{4 fig:RIS opt Power T vary mu} and \ref{4 fig:RIS grad abs Power T vary mu} show the convergence behaviors of the DCMC capacity and the absolute value of the gradient vector in the proposed algorithm, respectively, where the transmit power varying from $0$~dBm to $30$~dBm.
As seen in Fig.~\ref{4 fig:RIS opt Power T vary mu}, a high $\mu$ value significantly speeds up the convergence, which allows us to attain a high DCMC capacity given the number of iterations. In Fig.~\ref{4 fig:RIS grad abs Power T vary mu}, the convergence speed increased upon increasing the $\mu$ value. It is also observed that using higher $\mu$ significantly speeds up the dropping of the L2 norm. Hence, the appropriate parameter $\mu$ has to be selected to strike the balance between the achievable DCMC capacity and the convergence speed.
It is seen in Fig~\ref{4 fig:RIS grad abs Power T vary mu} that the gradient norm failed to reach a stationary point when an improper step size was used.
To be specific, for the transmit power from $P_{t} = 10$~dBm to $30$~dBm, the proposed algorithm with $\mu = 100$ exhibited the inconsistent transition, and for $P_{t} = 20$~dBm, the same fluctuation was induced even when the step size was as low as $\mu=10$.
Such unstable behaviors were typically observed in the GA-based algorithm when the step size is significantly high in the optimization problem considered, where the variables changed back and forth around the stationary point.
By contrast, the algorithm attained a proper convergence for $(P_{t}, \mu) = (30, 10)$ unlike for the case of $(P_{t}, \mu) = (20, 10)$.
Since discrete-valued inputs, i.e., $K$-sized constellation and $J$ RIS group patterns, are used in our system,  the achievable DCMC capacity is upper-bounded by $\log_2 K + \log_2 L$.
Therefore, when the transmit power is sufficiently high, many potential combinations of RIS phase shifts exist, which achieve the upper bound capacity. This implies that the upper bound of the DCMC capacity is achievable even with non-optimal beamforming with a high $\mu$ value.

\subsubsection{Information Rate Analysis}
\label{4 subsubsec:SM-Information Rate Analysis}
\begin{figure}
\centering
\includegraphics[width=.7\linewidth]{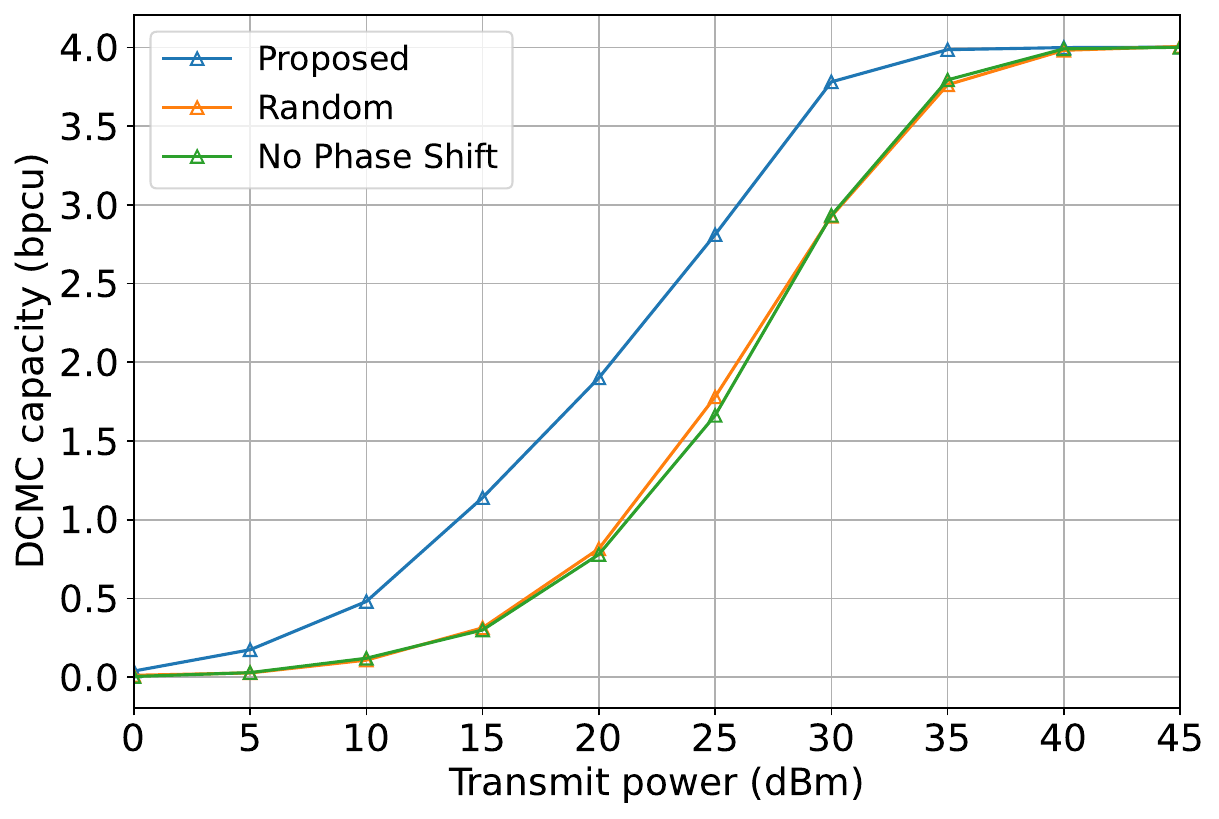}
\caption{The DCMC capacity of the proposed optimization scheme of our RIS-aided SM scheme.}
\label{4 fig:RIS mutual info vary T}
\end{figure}

Fig.~\ref{4 fig:RIS mutual info vary T} shows the DCMC capacity of the proposed optimization scheme in our RIS-aided SM scheme. The two conventional RIS schemes are selected as benchmarks, i.e., the random phase shifts and no phase shifts.
In the random phase shift scheme, the phase shifts of all the RIS groups are generated randomly from the uniform distribution [$0$, $2\pi$]~\cite{wu2019intelligent} in each symbol duration, while in the no phase shift scheme, all the phase shift coefficients are fixed to $\phi_{g} = 0$ throughout the transmission.~\cite{nagaya2024reduced}.
The data frame length of the CP was set as $(N_{b}, N_{CP}) = (1024, L_{0})$, and the DCMC capacity is calculated based on the BCJR algorithm~\cite{BCJR}.
In Fig.~\ref{4 fig:RIS mutual info vary T}, the proposed scheme outperforms the two benchmark schemes in the entire transmit power regime.
More specifically, the proposed scheme's $6$-dB advantage over the two benchmark schemes is attained for $1$ bpcu.

\subsubsection{Sensitivity Analysis}
\label{4 subsubsec:SM-Sensitivity Analysis}
\begin{figure}[!t]
\centering
\subfloat[]{\includegraphics[width=0.25\linewidth]{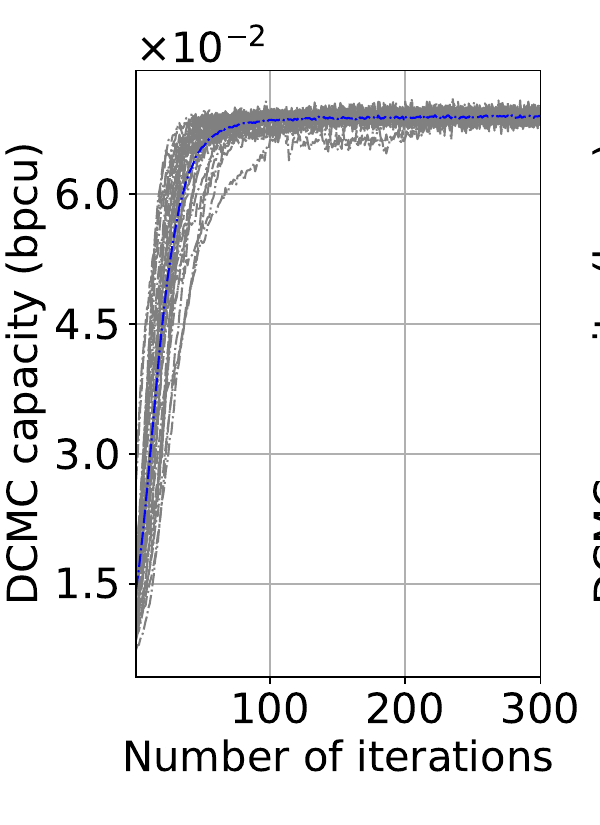}%
\label{4 fig:RIS sensitivity Power T 0dBm}}
\hfill
\subfloat[]{\includegraphics[width=0.25\linewidth]{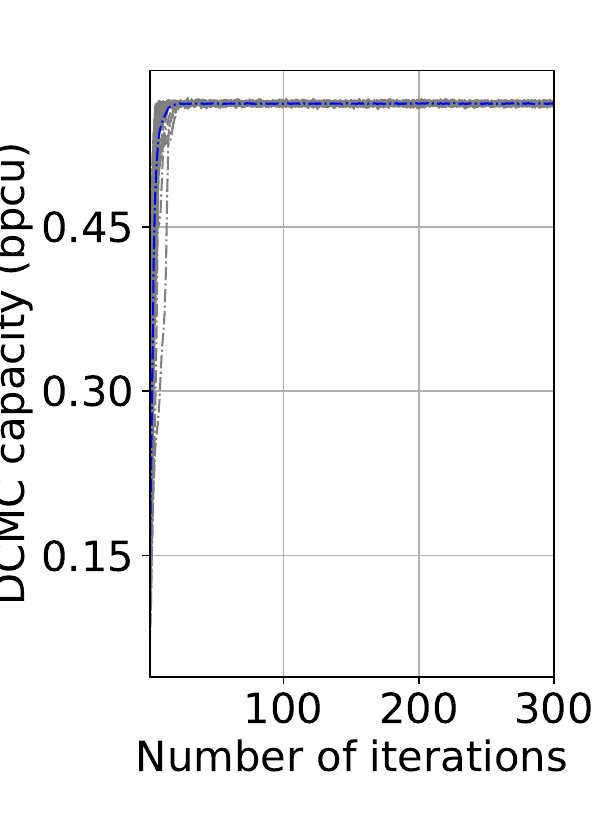}%
\label{4 fig:RIS sensitivity Power T 10dBm}}
\subfloat[]{\includegraphics[width=0.25\linewidth]{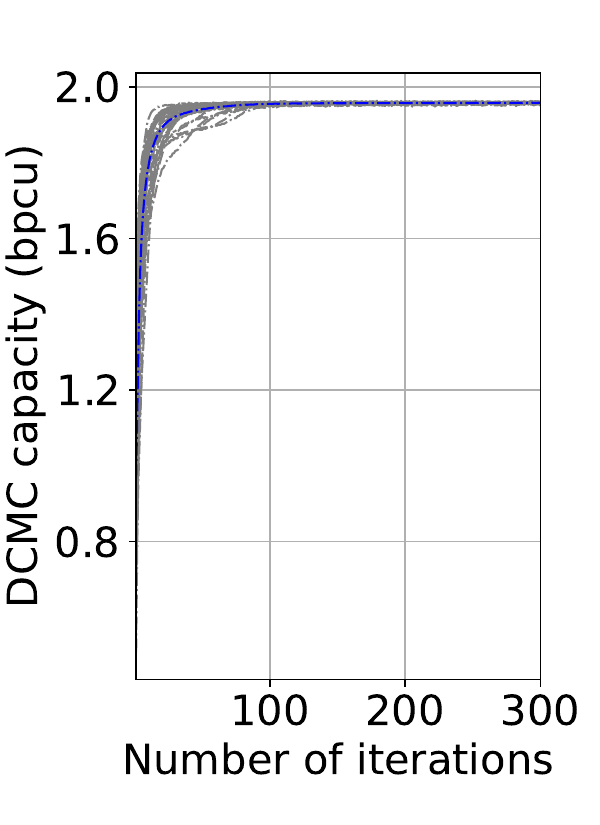}%
\label{4 fig:RIS sensitivity Power T 20dBm}}
\hfill
\subfloat[]{\includegraphics[width=0.25\linewidth]{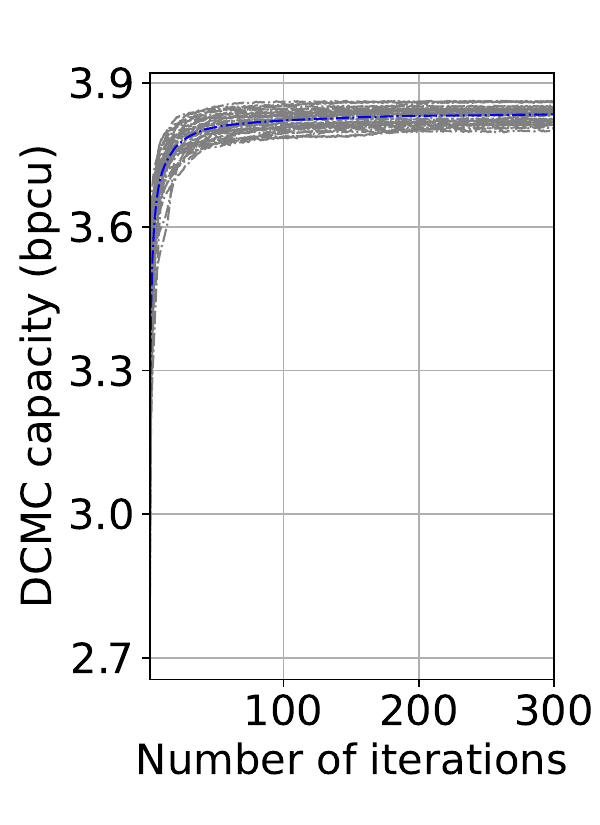}%
\label{4 fig:RIS sensitivity Power T 30dBm}}
\caption{Convergence analysis of the DCMC capacity with various initial conditions:
(a) $P_{t} = 0$~dBm, (b) $P_{t} = 10$~dBm, (c) $P_{t} = 20$~dBm, and (d) $P_{t} = 30$~dBm.}
\label{4 fig:RIS sensitivity}
\end{figure}

In Fig.~\ref{4 fig:RIS sensitivity}, we provide the sensitivity analysis for the convergence behavior of the proposed optimization algorithm depending on the initial values of the RIS phase shift vector. To precisely analyze the sensitivity, the step size was set as $\mu=10$ for $P_{t} = 0$, $10$~dBm and $\mu=1$ for $P_{t} = 20$, $30$~dBm, based on the convergence behavior of Figs.~\ref{4 fig:RIS opt Power T vary mu} and \ref{4 fig:RIS grad abs Power T vary mu}.
From Fig.~\ref{4 fig:RIS sensitivity}, it was found that
while the initial phase shift vector slightly affects the converged DCMC capacity, the DCMC capacity close to the average value drawn in the blue line was mostly attained. To circumvent convergence to a local optimum point with a low performance, it may be effective to adapt the step size $\mu$ in the optimization process when conversion is detected, as proposed in \cite{sugiura2021joint}.

\section{Conclusions}
\label{4 sec:Conclusions}
In this paper, we proposed novel RIS-aided single-carrier transmission schemes based on SM principle for the frequency-selective fading channel. By switching the activation and deactivation of the RIS groups for each transmit symbol, information bits are transferred simultaneously by the conventional PSK symbols and the SM patterns.
We derived the DCMC capacity of the proposed RIS-aided SM scheme in the frequency-selective fading channel, which is then used for our iterative RIS optimization algorithm as the cost function to be maximized.
To allow efficient optimization, we formulated the gradients of the DCMC capacity with respect to the RIS phase shift vector in the closed form.
In our performance results, it was demonstrated that the proposed schemes achieve better performance than the conventional benchmark schemes while analyzing the proposed optimization algorithm in a comprehensive manner.

\section*{Acknowledgement}
This work was supported in part by the Japan Science and Technology Agency (JST) ASPIRE (Grant JPMJAP2345), and in part by the Japan Society for the Promotion of Science (JSPS) KAKENHI (Grants 23H00470, 23K22752, 24K21615).
{
\bibliographystyle{IEEEtran}
\bibliography{IEEEabrv,bib}
}
\end{document}